\documentclass[10pt,letterpaper]{article}
\usepackage{opex3}

\usepackage{color}
\usepackage{graphicx}        
\usepackage{amsmath,amssymb}
\usepackage{dsfont}
\usepackage[isolatin]{inputenc}
\usepackage{caption}
\usepackage{subfig}

\DeclareMathOperator*{\argmin}{argmin}
\DeclareMathOperator*{\argmax}{argmax}

\newcommand{\ie}{\emph{i.e., }}

\newcommand{\Rr}{\mathds{R}}

\newcommand{\D}{\mathbf{D}}

\newcommand{\X}{\mathbf{X}}
\newcommand{\Y}{\mathbf{Y}}

\newcommand{\xv}{\mathbf{x}}

\newcommand{\yv}{\mathbf{y}}

\newcommand{\dv}{\mathbf{d}}

\newcommand{\rv}{\mathbf{r}}

\newcommand{\tv}{\text{\boldmath$\theta$}}

\newcommand{\Ns}{\mathcal{N}}

\newcommand{\Os}{\mathcal{O}}

\definecolor{orange}{rgb}{1,0.5,0}

\begin{document}

\title{Reference-less measurement of the transmission matrix of a highly scattering material using a DMD and phase retrieval techniques}

\author{Ang\'elique Dr\'emeau$^1$, Antoine Liutkus$^{2}$, David Martina$^{3}$, \\  Ori Katz$^{3,4}$, 
Christophe Sch\"ulke$^5$, Florent Krzakala$^{1,6}$, \\ Sylvain Gigan$^{3,6,*}$, Laurent Daudet$^{4,5}$  }

\address{$^1$LPS-ENS and CNRS UMR 8550, Paris, F-75005, France.\\
$^2$Inria, CNRS, Loria UMR 7503, Villers-lès-Nancy, F-54600, France\\
$^3$Laboratoire Kastler Brossel, Université Pierre et Marie Curie, Ecole Normale Supérieure, Collège de France, CNRS UMR 8552, Paris, F-75005, France \\
$^4$Institut Langevin, ESPCI and CNRS UMR 7587, Paris, F-75005, 
France\\
$^5$Paris Diderot University, Sorbonne Paris Cité, Paris, F-75013, France\\
$^6$Sorbonne Universit\'es, UPMC Universit\'e Paris 06, F-75005, Paris, France\\
}

\email{$^*$sylvain.gigan@lkb.ens.fr}

\begin{abstract}
This paper investigates experimental means of measuring the transmission matrix (TM) of a highly scattering medium, with the simplest optical setup.  Spatial light modulation is performed by a digital micromirror device (DMD), allowing high rates and high pixel counts but only binary amplitude modulation. We used intensity measurement
 only, thus avoiding the need for a reference beam. Therefore,  the phase of the TM has to be estimated through signal processing techniques of phase retrieval. Here, we compare four different phase retrieval principles on noisy experimental data. We validate our estimations of the TM on three criteria : quality of prediction, distribution of singular values, and quality of focusing. Results indicate that  Bayesian phase retrieval algorithms with variational approaches provide a good tradeoff  between  the computational complexity and the precision of the estimates. 
\end{abstract}


\bibliographystyle{osajnl}
\bibliography{Biblio,Biblio2}

\section{Introduction}

Wave propagation in complex media is a fundamental problem in physics, be it in acoustics, optics, or electromagnetism \cite{sebbah_waves_2001}. In optics, it is particularly relevant for imaging applications. Indeed, when light passes through a multiply scattering medium, such as a biological tissue or a layer of paint, ballistic light is rapidly attenuated, preventing conventional imaging techniques,  and random scattering events generate a so-called speckle pattern that is usually considered useless for imaging. Recently,  wavefront shaping using spatial light modulators (SLM) has emerged as a unique tool to manipulate multiply scattered coherent light, for focusing or imaging in scattering media \cite{mosk_controlling_2012}.  In essence, these methods use the linearity and time-reversal symmetry of the wave propagation, whatever the complexity of the medium, to control the output speckle field, by manipulating the light beam impinging on the scattering sample.  Different wavefront shaping approaches rely on digital phase-conjugation \cite{cui_implementation_2010, papadopoulos_focusing_2012} or iterative algorithms \cite{vellekoop_focusing_2007}, but it is also possible to measure the so-called transmission matrix (TM) of the medium \cite{popoff_measuring_2010}, which fully describes light propagation through the linear medium, from the modulator device to the detector. This approach has been particularly efficient for focusing, imaging \cite{popoff_image_2010, choi_overcoming_2011} and for studying the transmission modes of the medium \cite{kim_maximal_2012}. These methods are not only valid for scattering material but can also be applied to other complex transmission system, most notably multimode fibers, turning them into minimal footprint endoscopes \cite{choi_scanner-free_2012, papadopoulos_high-resolution_2013, bianchi_multi-mode_2012, cizmar_shaping_2011}. 

\begin{figure}[h!]
\begin{center}
\includegraphics[width=0.8\columnwidth]{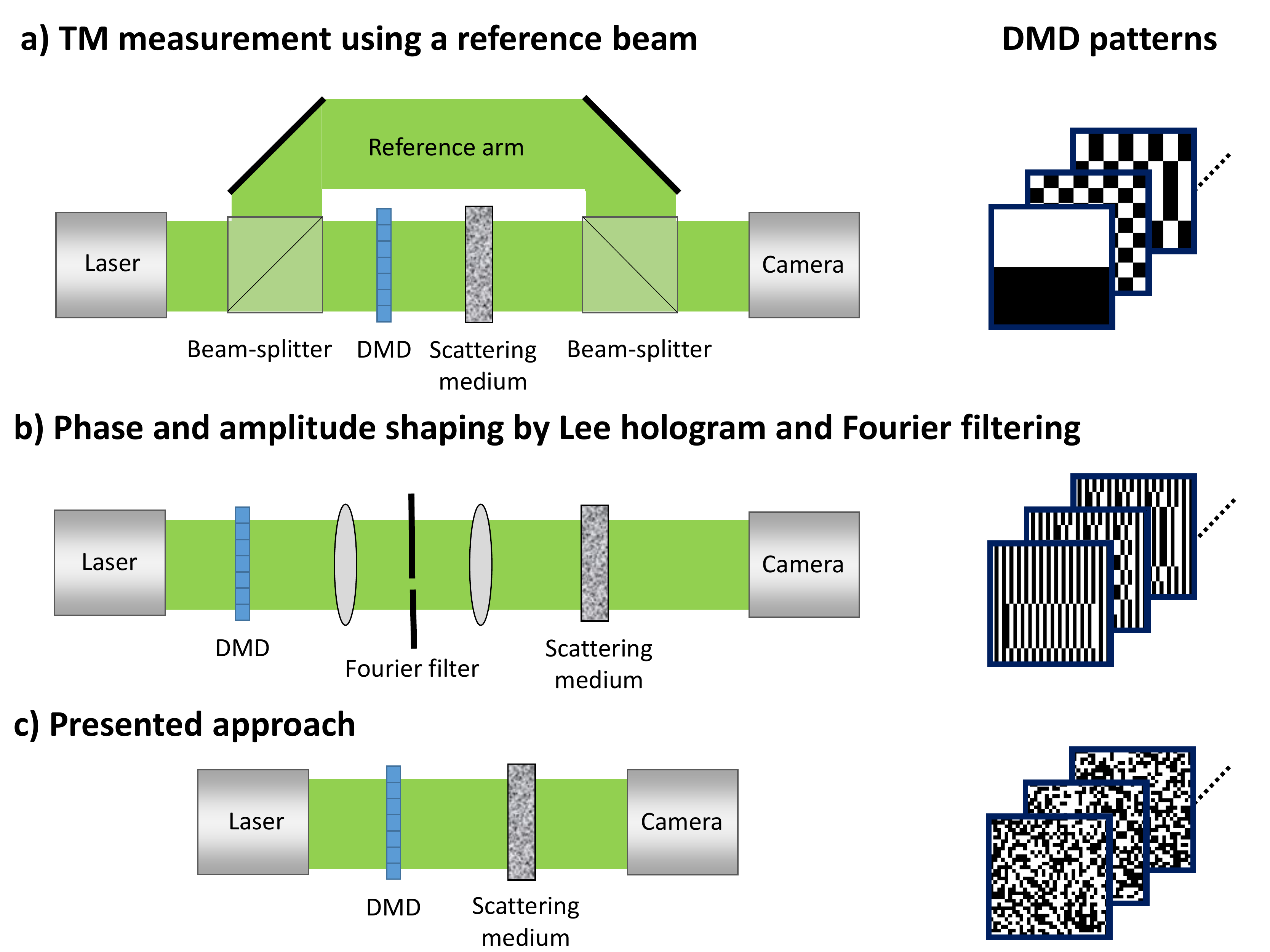}
\caption{Different experimental approaches for measuring the complex-valued transmission matrix of a scattering medium  with a binary DMD amplitude modulator. (a) use of a reference arm for retrieving the phase of the output field by off-axis or phase-shifting holography; (b) Using the DMD as a spatial phase modulator by displaying amplitude holograms, and using the unmodulated parts of the field as a phase-stable reference; (c) the presented approach where only intensity values are measured.}
\label{fig:measuringTM}
\end{center}
\end{figure}

A major limitation of most of these techniques for imaging is their speed. Indeed, the wavefront shaping process must be faster than the stability time of the medium, which can be of only a few milliseconds  in biological tissues. Yet, most of the works reported so far have relied on phase modulators which are usually slow (few tens of Hertz for liquid crystal modulators). Micro Electro-Mechanical Systems (MEMS) modulators are much faster, but are usually not phase-only. As a promising alternative for  wave shaping in complex media, Digital Micromirror Device (DMD) technology \cite{sampsell1995dmd} offers binary amplitude modulators (\ie ON or OFF) operating at $>20kHz$, with high pixel counts ($10^6$) and low pitch (around 10 microns), all this at  low cost. These binary amplitude modulators have been used as phase modulators, using appropriate diffraction and filtering, e.g. by Lee-type amplitude holography \cite{conkey_high-speed_2012,goorden_superpixel-based_2014}, as shown on Fig.~\ref{fig:measuringTM}b). While phase control is more effective for wavefront shaping than amplitude control, some works reported on using DMD as genuine binary amplitude modulators for wavefront shaping through opaque scattering media, albeit usually yielding lower overall efficiency than phase modulators for focusing or mode matching \cite{akbulut_focusing_2011, kim_implementing_2014, Tay:14}. 
The DMD configuration can also be optimized using genetic algorithms  \cite{kner_2014} to maximize the intensity  enhancement. 

For the measurement of a TM, an additional issue lies in  accessing the amplitude and phase of the output field, that in optics usually requires a holographic measurement, i.e. a reference beam, as shown on Fig.~\ref{fig:measuringTM}a). This reference beam can either be co-propagating in the medium \cite{popoff_image_2010, chaigne_controlling_2014}, or use an external reference arm \cite{akbulut_measurements_2013, choi_overcoming_2011}. The phase and amplitude of the measured field can then be extracted by simple linear combinations of interference patterns with a phase-shifted or off-axis reference. This however poses the unavoidable experimental problem of  the interferometric stability of the reference arm.

\medskip

In this work we report on the full measurement of the complex TM of a multiply scattering medium, using  a DMD binary amplitude modulator as an SLM, with no reference on the detection side, as shown on Fig.~\ref{fig:measuringTM}c). This approach  combines the high-speed and high pixel counts allowed by DMD devices, with the simplicity and robustness of a reference-less optical setup. However, it involves advanced signal processing algorithms for phase retrieval, run on a sufficiently large number of input-output calibration measurements.
In this study, we compare the performance of four phase-retrieval algorithms \cite{Gerchberg1972,Waldspurger2013,Schniter2012, Dremeau}, for the estimation of a TM based on actual noisy experimental measurements. We assess their performance as a function of the number of measurements, and compare their relative computational cost. 
We then show that the distribution of the singular values of the measured TM varies according to random matrix theory. 
Finally, we demonstrate that single- or multi-point light focusing can be achieved, using an $\ell_\infty$-regularization algorithm \cite{Fuchs2011} for determining the optimal DMD binary input pattern. In addition to being an interesting signal processing problem, our approach is particularly relevant for real-life applications of the TM approach, since it allows a simple, fast and robust implementation.

\section{Experimental setup} 

\begin{figure}[h!]
\begin{center}
\includegraphics[width=0.8\columnwidth]{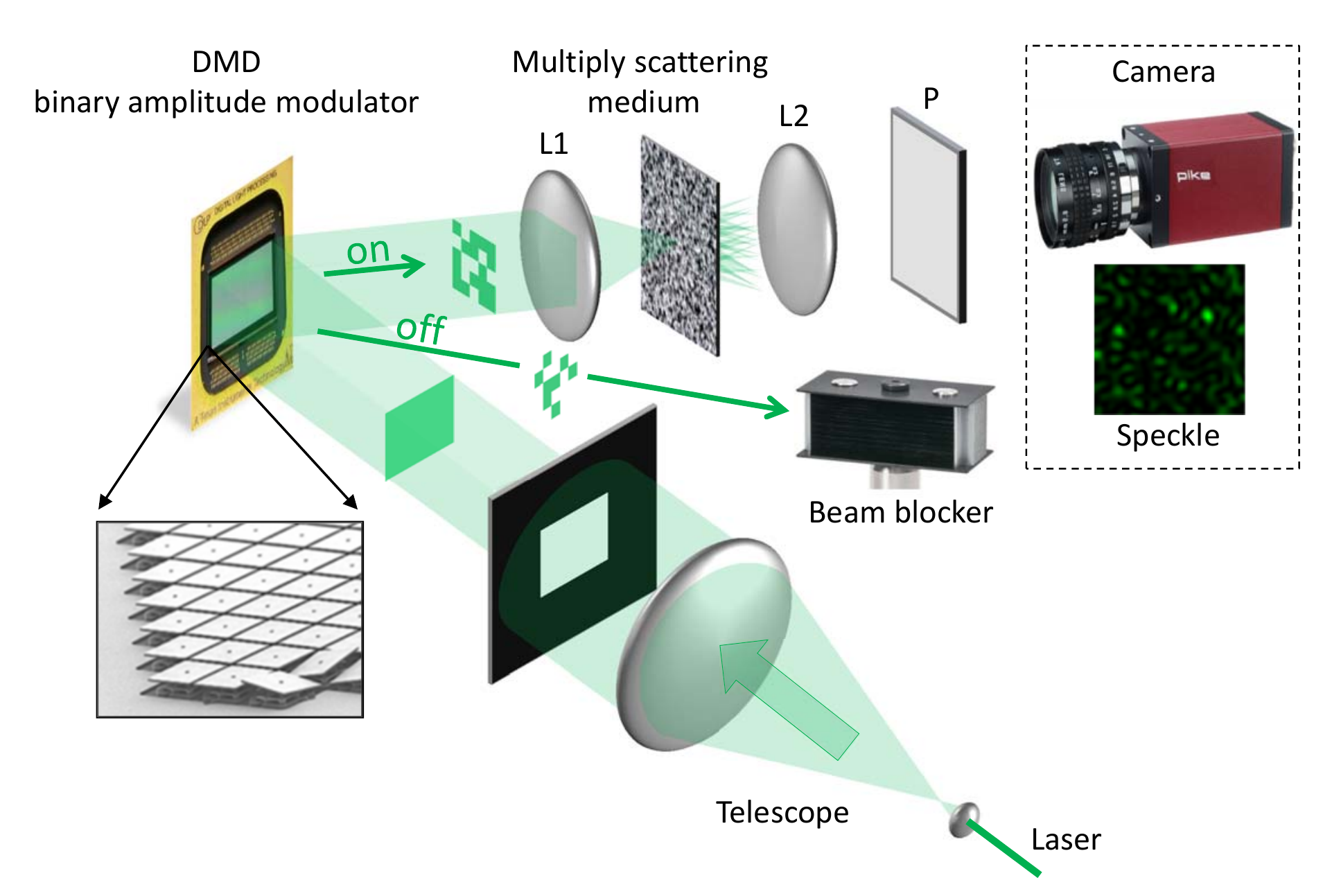}
\caption{Experimental scheme:  A $532$ nm CW laser is expanded through a telescope in order to obtain an homogeneous beam. Through a rectangular mask, it illuminates the DMD which acts as binary amplitude spatial light modulator. The DMD reflects the light in two different directions corresponding to either ON (unit transmission) or OFF (the light is deviated towards a beam dump). The transmitted pattern is focused by a first lens L1 on the scattering medium -- here a white paint layer --, acting as a thick multiply scattering medium. The transmitted speckle pattern is collected by a microscope objective and is observed through a polarizer P on a CCD camera.}
\label{fig:Exp_setup}
\end{center}
\end{figure}

Our experimental setup, described in Fig. \ref{fig:Exp_setup}, uses a DMD-array from Texas Instrument ($1920\times1080$ tilting micromirrors), driven by the DLP V-9500 VIS module (Vialux). The DMD is made of mirrors that can switch between two angular positions separated by $24^\circ$, thus reflecting each pixel either toward a beam dump (pixel OFF) or towards the focusing system (pixel ON).

Under Matlab, an amplitude mask is computed and loaded on the DMD. The pattern corresponding to the ON pixels is focused on the surface of a thick scattering medium by means of a $f=100$ mm lens L1 (thus the DMD pixels correspond roughly to incidence angles on the sample). The sample is a  $\sim 100$ microns thick layer of white paint, which is thick enough in order to considerably mix  the light on the other side, producing a complex speckle interfering pattern. This speckle pattern is collected through a microscope objective (L2) and detected on a camera (AVT Pike F-100B).
In order to measure the TM, we need to send a large series of input patterns (typically a few times the number of input pixels we wish to control), in a time over which the medium can be considered stationary.  
For this purpose,  we use the ``high speed" driver provided with the DMD in order to load all the to-be-projected random amplitude masks to the memory of the DMD driver module, and we trigger the display of each mask via a DAQ card (National Instruments, PCI-6221) and a waveform generator. 
In the same way, in order to be as fast as possible, we also only consider a subregion  on the camera of  size of $400\times400$ pixels.  The overall acquisition rate is  $31$ images/second. To monitor the stability of the medium, we periodically measure the correlation of the speckle image corresponding to the same input mask. We therefore quantify the stability of the medium, which is better than $98$\% over the total measurement time (typically around $5$ minutes). 

\section{Estimating the TM with intensity-only measurements and binary inputs}\label{sec:pr}

The measurement of the TM can be formalized as a calibration problem:
given $P$  incoming waves, assumed perfectly known, which model explains at best the observed outputs? In our case, this inverse problem reduces to the well-known problem of phase retrieval.

Let $\xv_\mu \in\lbrace 0, 1 \rbrace^N$ stand for the binary DMD inputs related to the $\mu$-th acquisition, where $N$ is the number of pixels (mirrors) used on the DMD. We assume that the partial observations of the sole moduli of the transmitted waves  (the square root of the camera measured intensities),
denoted by $\yv_\mu \in\Rr_+^M$, obey
 \begin{align}
 \yv_\mu = |\D\xv_\mu|, \quad\quad\quad\forall \mu\in\lbrace1,\ldots,P\rbrace,\label{eq:model} 
 \end{align}
where $\D$ is the TM complex-valued transmission matrix characterizing the scattering material, and $M$ is the number of observed pixels on the camera.

Then, adopting a matrix formulation and conjugating-transposing the system, we get 
\begin{align}
\Y^H = |\X^H\D^H|,
\end{align}
where $\Y=[\yv_1,\ldots,\yv_P]$, $\X=[\xv_1,\ldots,\xv_P]$  and $.^H$ denotes the conjugate-transpose of a matrix/vector. This reveals a ``classic" phase retrieval problem:
 given the matrix of inputs $\X^H$, each column of $\Y^H$ is used to estimate each complex-valued column of $\D^H$.

\subsection{Phase retrieval}
The problem of reconstructing a complex vector given only the magnitude of measurements 
is a non-convex optimization problem notoriously difficult to solve. Many algorithms have been devised in the literature to deal with this problem. We can roughly divide them into three main families:

\begin{enumerate}
\item \emph{The alternating-projection algorithms} alternate projections on the span of the measurement matrix and on the object domain. Among these approaches, we can mention the works of Gerchberg \& Saxton \cite{Gerchberg1972}, Fienup \cite{Fienup1982} and Griffin \& Lim \cite{Griffin1984}. 
\item \emph{The algorithms based on convex relaxations} approximate the phase recovery problem by relaxed problems which can be solved efficiently by standard optimization procedures. Two of the main approaches of this type, namely \emph{PhaseLift} \cite{Candes2013} and \emph{PhaseCut} \cite{Waldspurger2013}, rely in particular on semidefinite programming. 
\item \emph{The Bayesian approaches}, recently envisaged in \cite{Schniter2012, Dremeau}, circumvent the non-linearity of the modulus through the introduction of hidden variables and resort to variational approximations to approximate the posterior distribution of the variables of interest. These latter methods have been shown to perform good reconstruction in a reasonable computational time \cite{Dremeau}. 
\end{enumerate}

\subsection{Bayesian variational approximations\label{subsec:calib}}
Additionally to the previous notations, we introduce new variables, modeling, on the one hand, the missing phases of the observations, and on the other hand, some acquisition noise. Thus, recalling that we resort to a conjugate-transposition of the matrix system, each absolute-valued measurement
$y_\mu$, $\mu\in\lbrace 1\ldots P\rbrace$, of any row $\yv$ of $\Y$, is expressed as
\begin{align}
y_\mu =e^{j\theta_\mu} \big(\sum_{i=1}^N x_{\mu i}\; d^{*}_{i} +\omega_\mu\big),\label{eq:y}
\end{align}
where $\theta_\mu\in[0,2\pi)$ stands for its missing conjugate phase, $x_{\mu i}$ is the $i$th element of the $\mu$th row in $\X$, $d^{*}_{i}$ corresponds to the $i$th conjugate element in the current estimated row $\dv$ of $\D$ and $\omega_\mu$ is an additive noise, assumed centered isotropic Gaussian
 (denoted $\mathcal{CN}$ in the following) with variance $\sigma^2$. 
We moreover suppose that the probability distributions for the entries of the matrix and for the missing phases are:
\begin{align}
&p(\dv) = \prod_{i=1}^N p(d_i) \quad\quad \text{with} \quad\quad p(d_i)=\mathcal{C}\Ns(0,\sigma_d^2),\label{eq:x}\\
\text{and} \quad \quad &p(\tv) = \prod_{\mu=1}^P p(\theta_\mu) \quad\quad \text{with} \quad\quad  p(\theta_\mu) = \frac{1}{2\pi}.\label{eq:th}
\end{align}
Under these assumptions, the absence of phases in the observations is naturally taken into account in the model since marginalizing on $\theta_\mu$ leads to a distribution on $y_\mu$ which only depends on the moduli of $y_\mu$ and $\sum_{i=1}^N x_{\mu i}\; d^{*}_{i}$.

Within model \eqref{eq:y}-\eqref{eq:th}, the recovery of the complex signal $\dv$ can be expressed as the solution of the following marginalized Maximum A Posteriori (MAP) estimation problem
\begin{align}
\hat \dv = \argmax_{\dv} p(\dv|\yv) ,\label{eq:pb}\\
\text{with} \quad\quad p(\dv|\yv) = \int_\tv p(\dv,\tv|\yv).
\end{align}
Because of the marginalization on the hidden variables $\tv$, the direct computation of $p(\dv|\yv)$ is however intractable in general. The solutions in \cite{Schniter2012, Dremeau} optimally approximate, in a Kullback-Leibler sense, the posterior joint distribution $p(\dv,\tv|\yv)$ by $\hat{q}(\dv,\tv)$  conditionally to a set of given constraints $\mathcal{F}$:
\begin{align}
\hat{q}(\dv,\tv) &= \argmin_{q\in\mathcal{F}} \int_\dv\int_\tv q(\dv,\tv) \log \big(\frac{q(\dv,\tv)}{p(\dv,\tv|\yv)} \big) d\dv \; d\tv. \label{eq:KL}
\end{align}
Depending on $\mathcal{F}$, the minimization \eqref{eq:KL} gives rise to different approximations. 
\begin{itemize}
\item In particular, $\mathcal{F} = \big\lbrace q\big| q=\prod_{i=1}^N q_i(d_i)\prod_{\mu=1}^P q_\mu(\theta_\mu)\big\rbrace$ defines a Mean-Field approximation, and problem \eqref{eq:KL} can be efficiently solved using the ``Variational Bayes Expectation-Maximization" (VBEM) algorithm \cite{Beal2003}. This is the approach considered in \cite{Dremeau}, denoted by \emph{prVBEM} in the rest of this paper.
\item With $\mathcal{F} = \big\lbrace q\big| q=\frac{\prod_{a=1}^A q_a(\dv_a)\prod_{b=1}^B q_b(\tv_b)}{\prod_{i=1}^N q_i(d_i)^{\alpha_i-1}\prod_{\mu=1}^P q_\mu(\theta_\mu)^{\beta_\mu-1}}\big\rbrace$ where $[\dv_1\ldots \dv_A]$ (resp. $[\tv_1\ldots \tv_B]$) partitions the variables $\dv$ (resp. $\tv$), and $\alpha_i$ (resp. $\beta_\mu$) is the degree of variable node $d_i$ (resp. $\theta_\mu$), problem \eqref{eq:KL} refers to the minimization of the Bethe free energy, which can be solved by
generalized approximate message passing (GAMP) algorithms, see \cite{Rangan2011}. This is the approach followed in \cite{Schniter2012}, denoted by  \emph{prGAMP} in the rest of this paper.
\end{itemize}
We will not detail here the structure of the resulting algorithms. We refer the interested reader to the papers \cite{Dremeau} and \cite{Schniter2012} and the authors' webpage\footnote{http://angelique.dremeau.free.fr/ (released October 7th, 2014)}
for a practical implementation of the \emph{prVBEM} algorithm. 

\subsection{Experiments and results}

\subsubsection{Prediction performance}
To assess the accuracy of the TM estimated by the considered approaches, we adopt a cross-validation-like experimental framework. The setup is as follows. We measure the $M=40000$ camera pixels stemming from $N=900$ DMD mirrors, $50$\% of them  being turned on, the others off  at each displayed pattern. The operation is repeated randomly $P=6000$ times. Given this dataset, a row of the TM is then learned from $p=\alpha N$ calibration measurements, with $\alpha$ varying in $\lbrace1,\ldots, 6\rbrace$, and used in a second step to predict the $P-p$ remaining measurements. This estimation is performed on $50$ different rows of the TM.

We evaluate and compare the performance of $4$ different algorithms: \emph{Gerchberg-Saxton} \cite{Gerchberg1972}, \emph{PhaseCut} \cite{Waldspurger2013}, \emph{prGAMP} \cite{Schniter2012} and \emph{prVBEM} \cite{Dremeau}. The algorithms present different complexities. The implementation of \emph{PhaseCut} (available on author's webpage\footnote{http://www.di.ens.fr/data/software/}) relies on interior-point methods, with a complexity growing as $\Os(p^{3.5}\log(1/\epsilon))$ where $\epsilon$ is the target precision \cite{Nemirovski2001}. \emph{Gerchberg-Saxton}, \emph{prGAMP} (in our own implementations) and \emph{prVBEM} share similar complexities, of order $\Os(p^2)$.
\begin{figure}[h!]
\begin{center}
	\subfloat[]{\includegraphics[width=0.47\columnwidth]{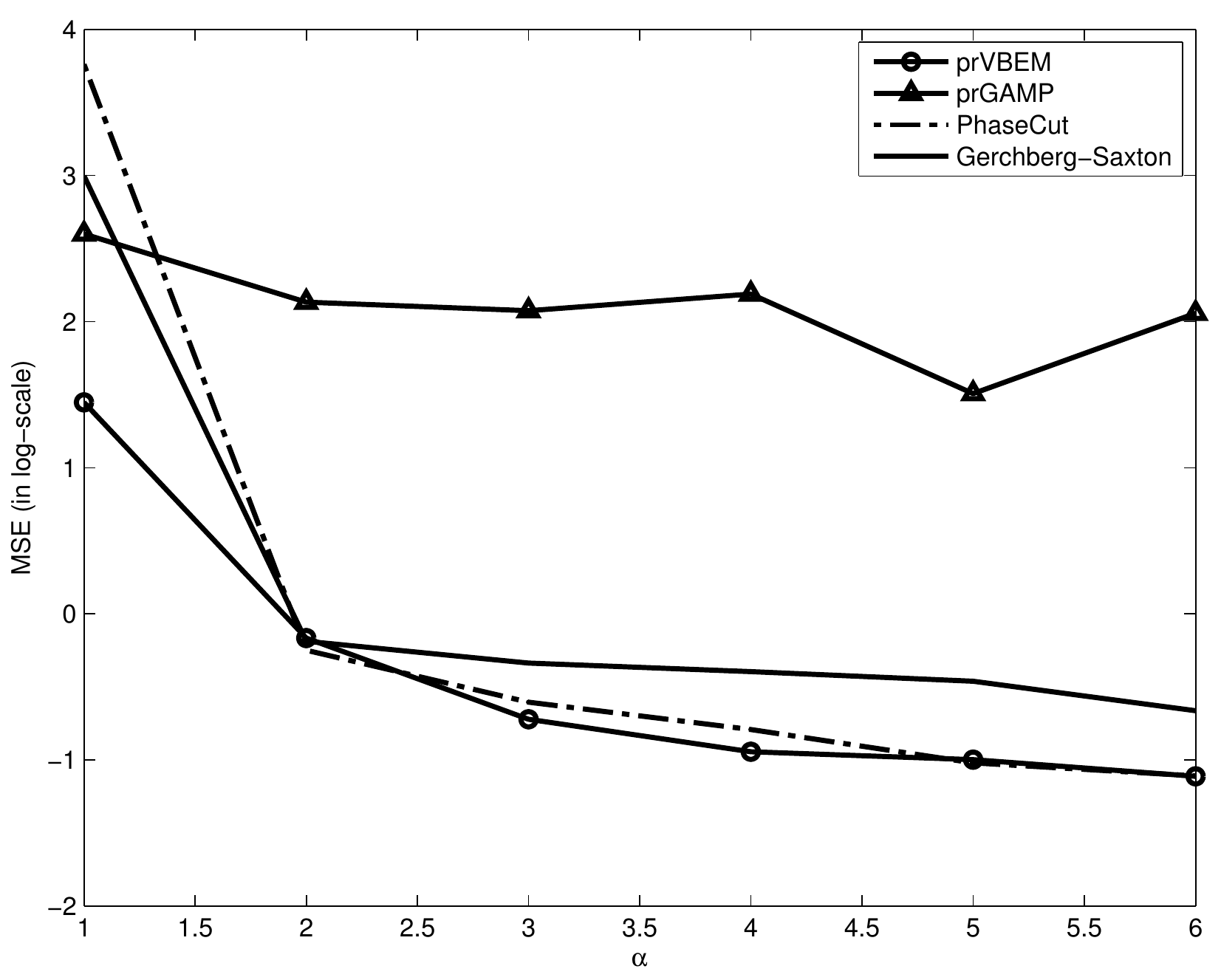}\label{fig:mse}}
	\subfloat[]{\includegraphics[width=0.48\columnwidth]{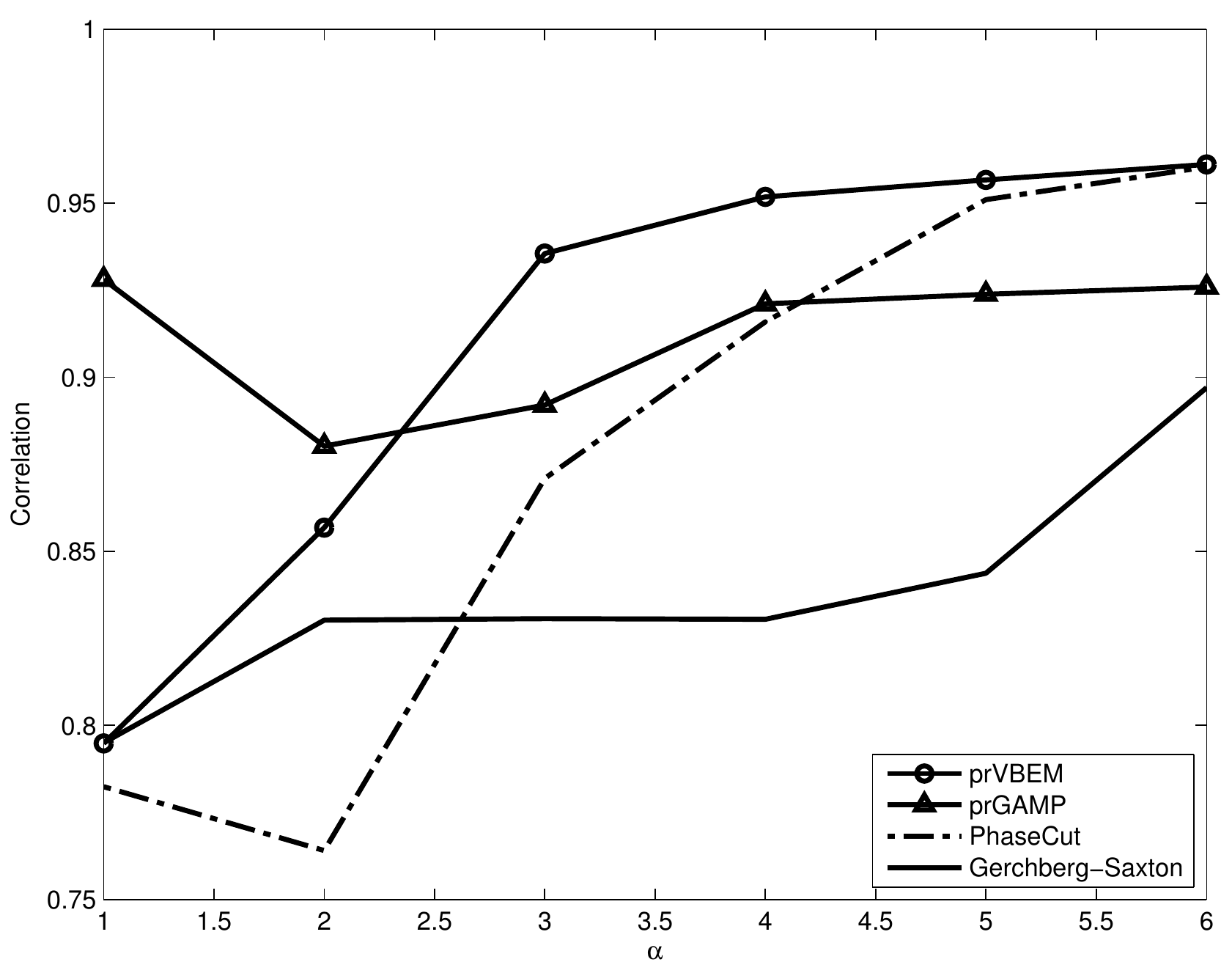}\label{fig:corr}}
\caption{Prediction performance according to (a), the mean-square error (MSE), in log scale, and to (b), the normalized cross-correlation between observation predictions using the estimated TM, and actual measurements of the output moduli (square root of the camera intensity values), as a function of the number of calibration measurements (x-axis is $\alpha$, such that $p=\alpha N$ calibration measurements are used). }
\end{center}
\end{figure}

\begin{figure}[h!]
\begin{center}
  \includegraphics[width=0.58\columnwidth]{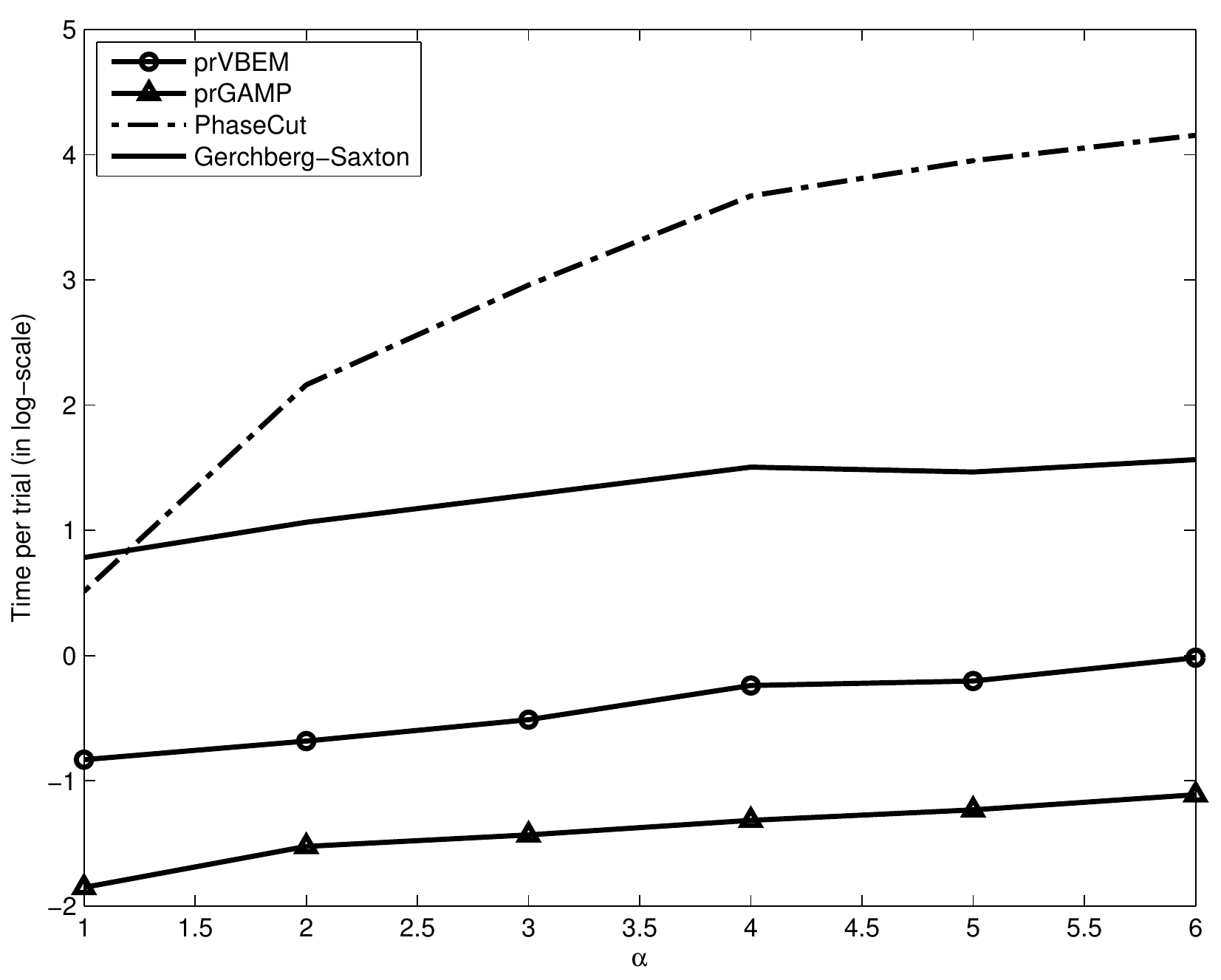} \caption{Average running time (log-scale, in seconds) as a function of the number of calibration measurements (x-axis is $\alpha$, such that $p=\alpha N$ calibration measurements are used). These simulations have been done on a Macbook Air with a 1.7GHz i7 processor.}\label{fig:tcrop} \end{center} \end{figure}

As a tradeoff between computational cost and performance, we set the stopping criteria for each algorithm  as follows. \emph{PhaseCut} is run until the target precision drops below $10^{-2}$.  \emph{Gerchberg-Saxton} stops after $3000$ iterations, \emph{prGAMP} and \emph{prVBEM} after $200$ iterations.	 Given the complexities of the algorithms, we allow \emph{PhaseCut} and \emph{Gerchberg-Saxon} for a higher running time, as shown and further discussed in Fig. \ref{fig:tcrop}.

The prediction performance of the algorithms is evaluated according to the mean-square error (MSE) (Fig. \ref{fig:mse})  and the normalized cross-correlation (Fig. \ref{fig:corr}) between the moduli of the $P-p$ predicted measurements and the actual observed ones. For \emph{Gerchberg-Saxton}, \emph{PhaseCut} and \emph{prVBEM}, the MSE curves present similar behaviors : they decrease monotonically with increasing $\alpha$. This observation resonates in Fig. \ref{fig:corr}, with the general increase tendency of the correlation. Interestingly, we see that for $\alpha\geq3$, that is, for at least $3$ times more real measurements than complex unknowns, \emph{prVBEM} outperforms all other algorithms with a correlation around 0.95.

On the contrary, \emph{prGAMP} presents a contrasted performance. If it leads to a good correlation (Fig. \ref{fig:corr}) between estimated and observed measurements, its MSE remains very high, independently of $\alpha$ (Fig. \ref{fig:mse}). Here, the algorithm finds an acceptable solution but only up to a multiplicative factor. While this is not an issue for the focusing experiment considered in Section \ref{sec:focus}, this could become a limitation in other more complex tasks.

Finally, as previously mentioned, 
we allow in these experiments more iterations for \emph{PhaseCut} and \emph{Gerchberg-Saxton}. 
In parallel to the performance curves exposed above, Fig. \ref{fig:tcrop} illustrates the corresponding average running time of the $4$ considered algorithms. In this figure, we see that \emph{prGAMP} performs the lowest computational cost, closely followed by \emph{prVBEM}. \emph{PhaseCut} requires a long running time, prohibitive in our application context. It should be noted that the \emph{Gerchberg-Saxton} algorithm, although relatively slow, still exhibits good performance, especially given its simplicity.  

On the basis of these preliminary experiments, for the remaining of this paper, we choose the \emph{prVBEM} approach and set the number of calibration measurements to $p=4N$, as a good tradeoff between performance and computation time. 
In this case, computing one row of the TM takes about .6 s in Matlab, on a Macbook Air with a 1.7GHz i7 processor, keeping in mind that rows are independent. 

\subsubsection{Comparison of singular values to Random matrix theory}
Interestingly, we can check that the measured TM presents some characteristics as predicted by random matrix theory. One practical way is to verify 
 that the distribution of its normalized singular values obeys the Mar\u{c}enko-Pastur law \cite{Marcenko1967}. 
 It should be noted that such apparently random signals are the hardest case for phase retrieval, where no specific structure can be taken into account.

\begin{figure}[h!]
\begin{center}
	\captionsetup[subfigure]{labelformat=empty}
	\subfloat[$\gamma=1$]{\includegraphics[width=0.3\columnwidth]{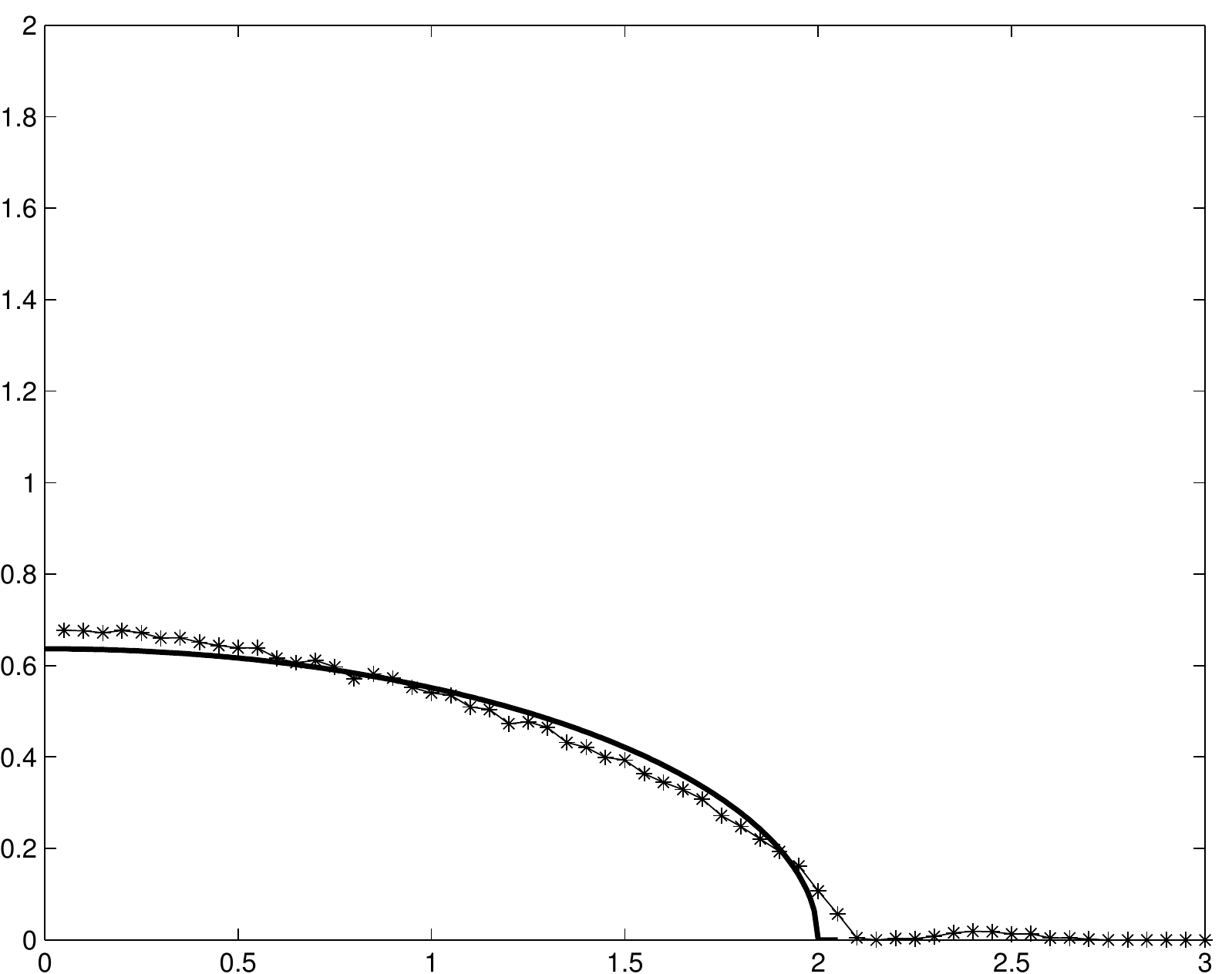}\label{fig:SV1}}\hspace{0.1cm}
	\subfloat[$\gamma=2$]{\includegraphics[width=0.3\columnwidth]{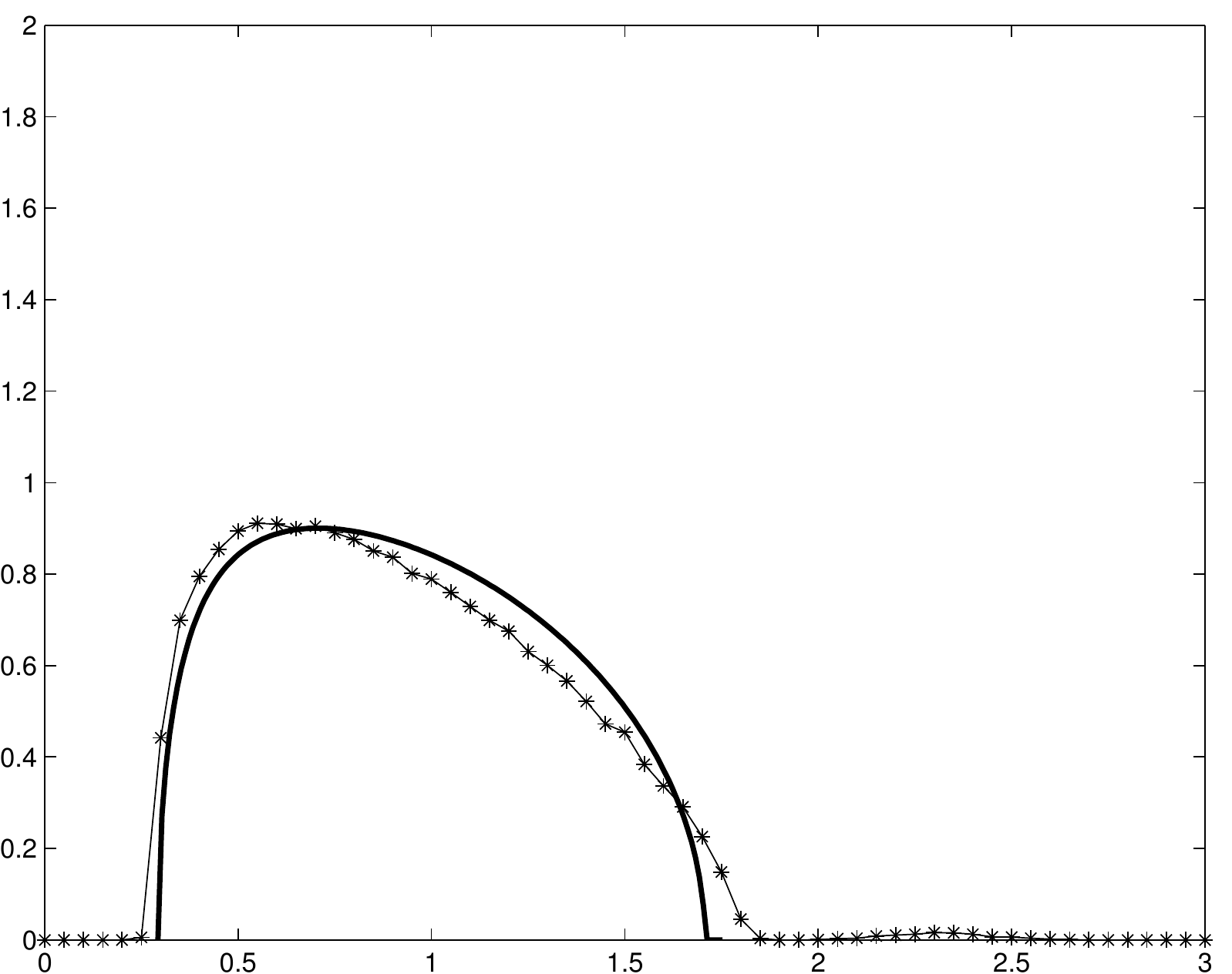}\label{fig:SV2}}\hspace{0.1cm}
	\subfloat[$\gamma=3$]{\includegraphics[width=0.3\columnwidth]{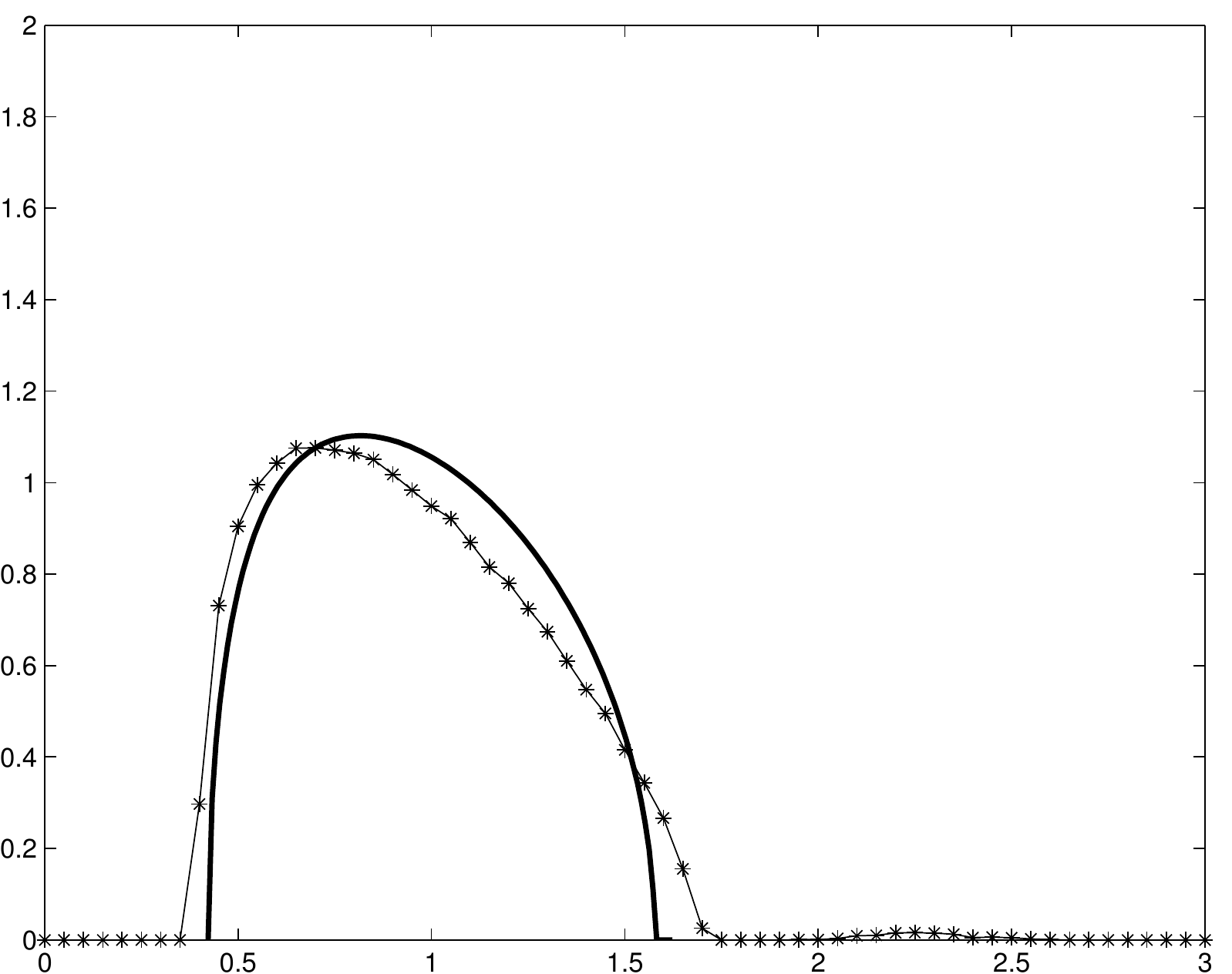}\label{fig:SV3}}\\
	\subfloat[$\gamma=4$]{\includegraphics[width=0.3\columnwidth]{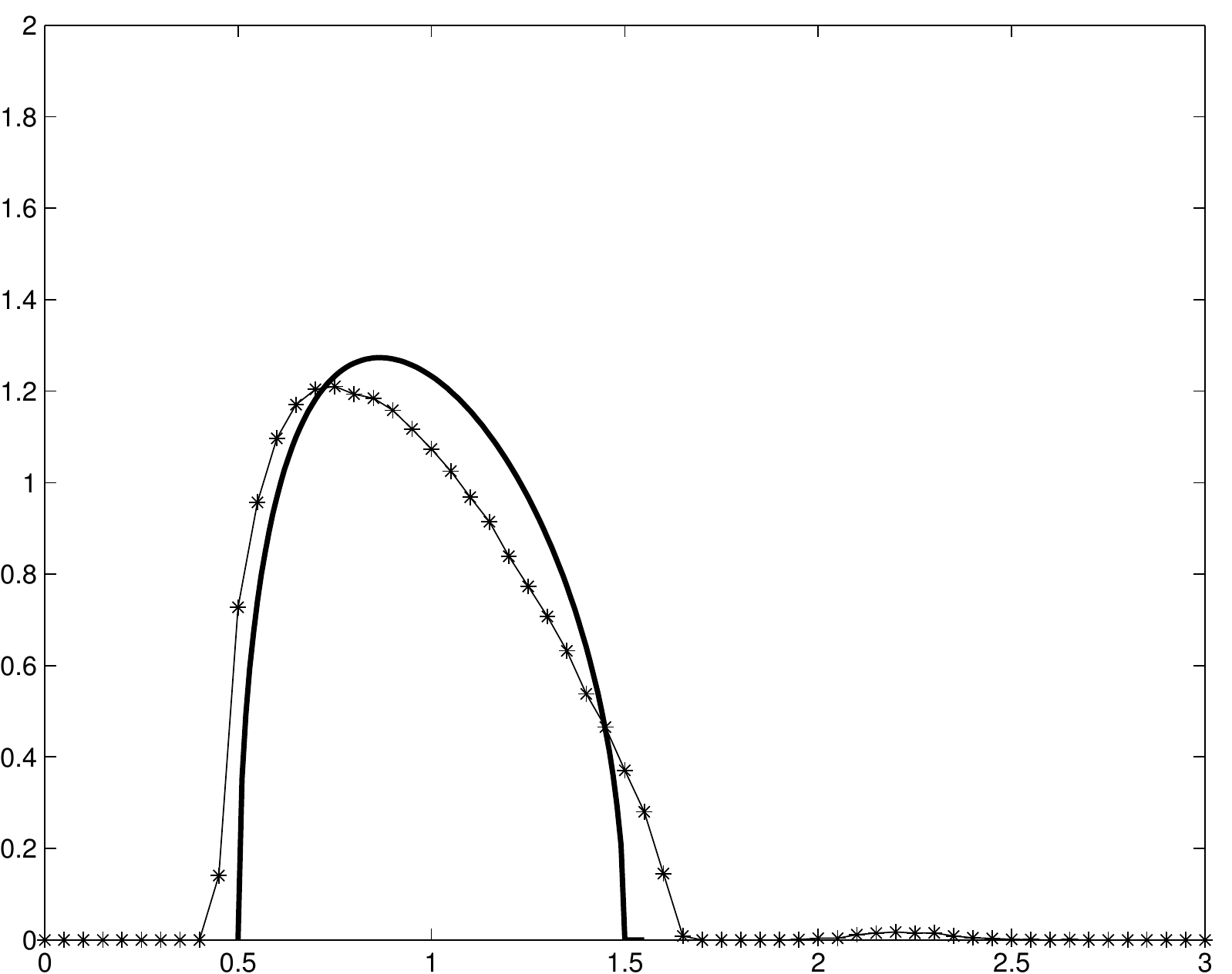}\label{fig:SV4}}\hspace{0.1cm}
	\subfloat[$\gamma=5$]{\includegraphics[width=0.3\columnwidth]{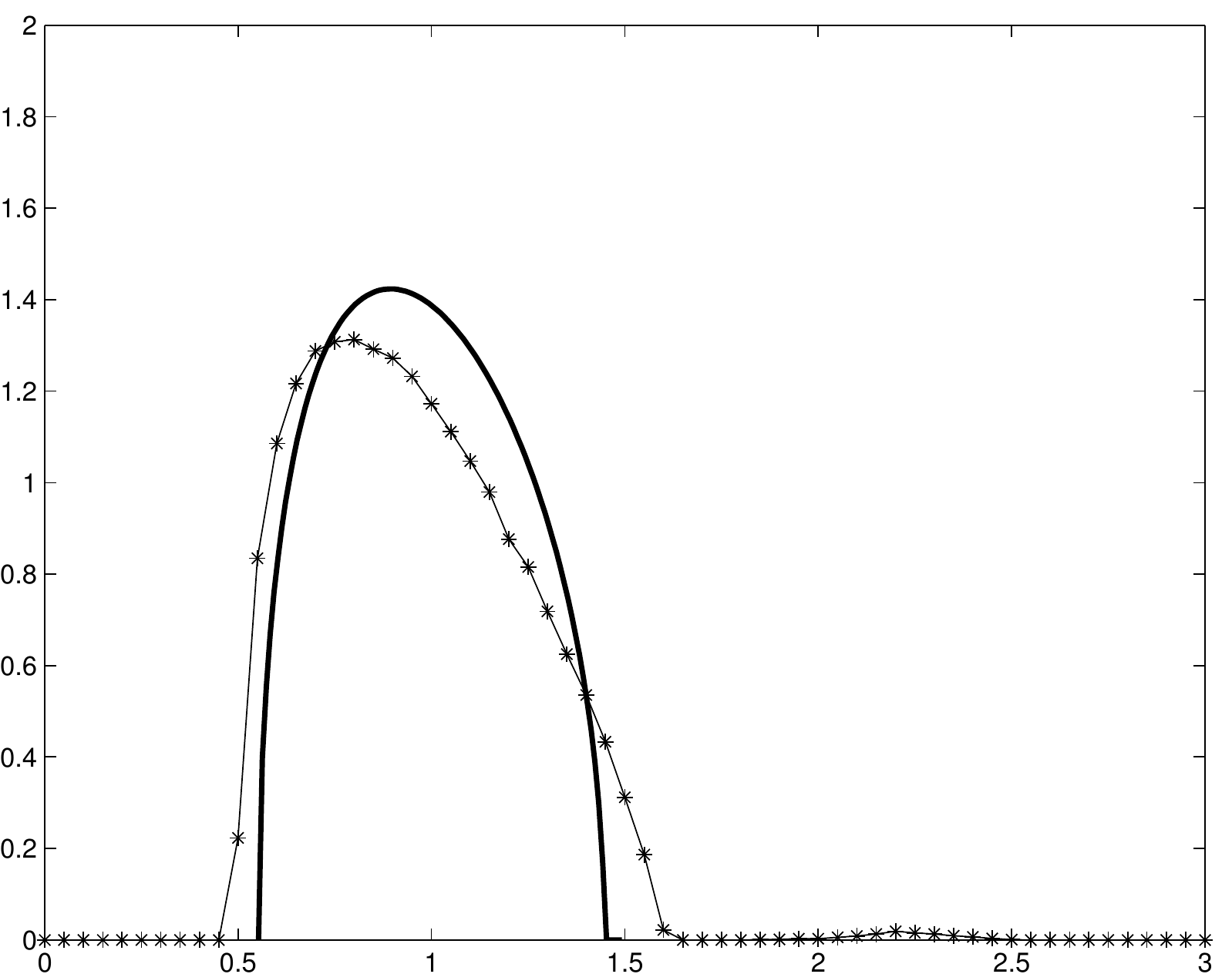}\label{fig:SV5}}\hspace{0.1cm}
	\subfloat[$\gamma=6$]{\includegraphics[width=0.3\columnwidth]{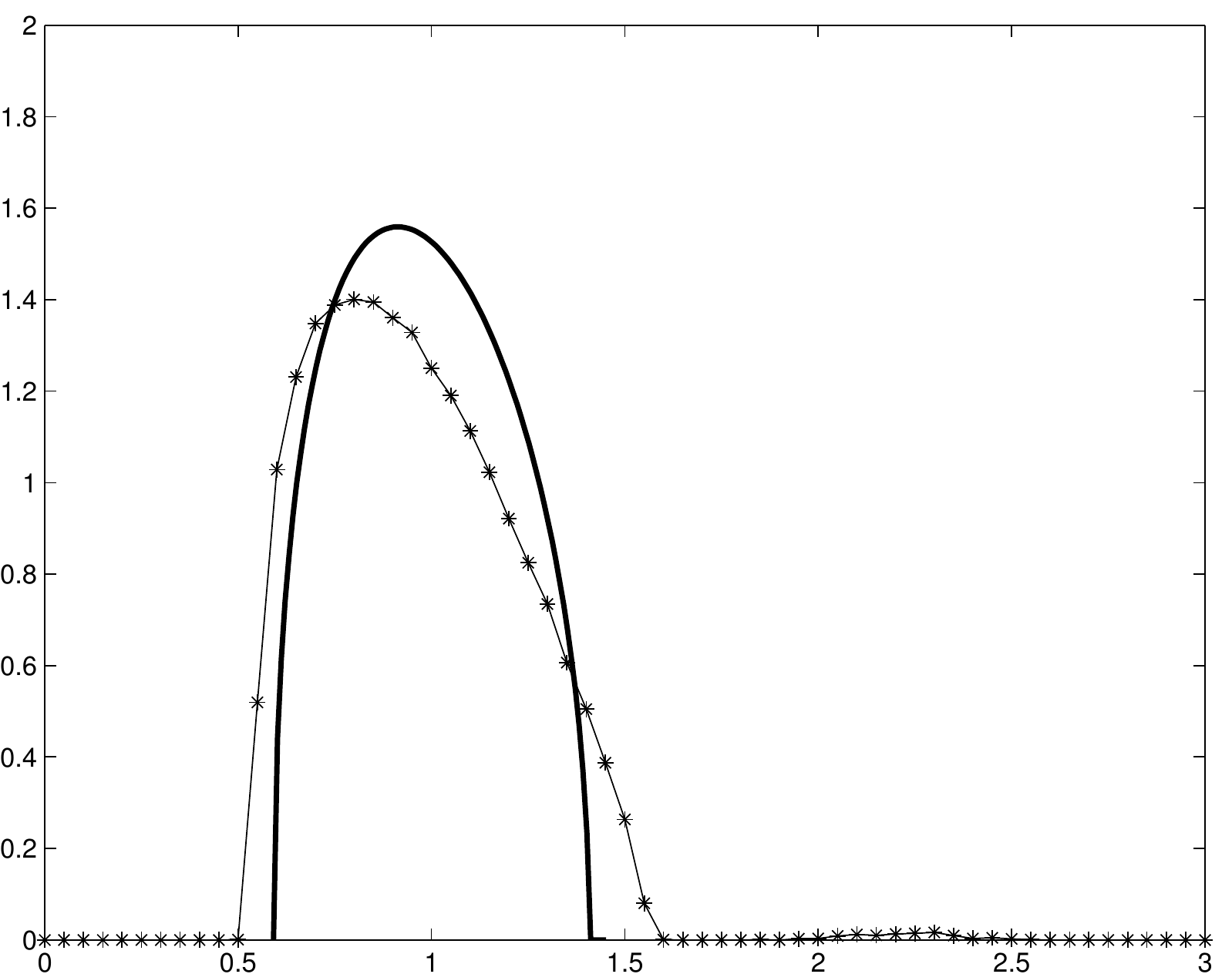}\label{fig:SV6}}
	\caption{Density of the normalized singular values for different $\gamma = M/N$. Stamped line: experimental results, continuous line: Mar\u{c}enko-Pastur law.}\label{fig:SV}
\end{center}
\end{figure}

In order to reduce the influence of specifics of our experimental setting, we perform the following operations, as in \cite{popoffNJP2011}:
\begin{itemize}
\item[\emph{i)}]  We normalize over the rows and columns, to attenuate the illumination artifacts: residual illumination ``by default'' on each pixel of the camera for the rows, and inhomogeneous contribution of each DMD mirror on the entire set of camera pixels for the columns.
\item[\emph{ii)}]  Because of the size of the speckle grains, two neighboring DMD mirrors may affect the material in the same way, as well, two pixels of the camera will be potentially correlated. To avoid this effect, we subsample  the rows and columns of the matrix.
\end{itemize}
To draw the empirical spectral density, we then consider the following setup.  We subsample the columns of the matrix up to $N=200$ and leave the number of rows varying, more precisely $M=\gamma N$, with $\gamma \in \lbrace 1,\ldots,6\rbrace$. These sub-matrices thus constitute partitions of the estimated matrix, randomly picked $100$ times to average the resulting densities. Fig. \ref{fig:SV} compares the experimental curves to the theoretical ones drawn according to the Mar\u{c}enko-Pastur law. We see that the experiments qualitatively follow the predictions. We remark, however that the larger $\gamma$ is, the more chances we have to consider the contributions of neighboring correlated pixels. This partly explains the increasing gap between both curves.

\section{Focusing with the DMD} 
\label{sec:focus}

Knowing the TM gives a powerful and flexible tool to control light within the scattering medium \cite{popoffNJP2011}. In particular, it can be used to 
compute which DMD input has to be set, in order to display a given arbitrary pattern at the receiver end. 
In this section, we demonstrate the special case of focusing light 
with maximum intensity on a desired pattern (a chosen sparse subset of the output pixels), with the TM measured experimentally as in the section \ref{sec:pr}. 
It should be emphasized that we keep the same 
experimental setup, with the binary DMD as  input device. Here, simple inversion methods such as \cite{popoffNJP2011} cannot be used, as these require intensity- or phase-modulated inputs. 

We propose here to resort to a similar Bayesian variational approach as for the calibration, adapted to the binary nature of the DMD inputs.

\subsection{Mean-Field-based inversion}
Formally, the problem can be expressed as an inverse problem, where, knowing the TM $\D$ and the observation $\yv$, we look for the DMD input $\xv$ such as described in \eqref{eq:model}. 
Adopting a similar modeling as in previous section, we then assume, for all elements $y_\mu$ with $\mu\in\lbrace1,\ldots,M\rbrace$,
\begin{align}
y_\mu =e^{j\theta_\mu} \big(\sum_{i=1}^N d_{\mu i}\; x_{i} +\omega_\mu\big),\label{eq:y_foc}
\end{align}
where $\theta_\mu\in[0,2\pi)$ stands for the missing conjugate phase, $d_{\mu i}$ is the $\mu$th element of the $i$th-column in $\D$, $x_i\in \{0, 1\}$ corresponds to the state of the $i$-th DMD pixel and $\omega_\mu$ is an additive noise, assumed centered isotropic Gaussian of variance $\sigma^2$. As in the subsection \ref{subsec:calib}, we suppose that the elements $\theta_\mu$ are independently and uniformly distributed in the interval $[0,2\pi)$, however, in order to accommodate for binary inputs, we consider here a Bernoulli model for $\xv$:
\begin{align}
p(\xv) = \prod_{i=1}^N p(x_i) \quad\quad \text{with} \quad\quad p(x_i) = \text{Ber}(p_i) =\left\lbrace 
	\begin{array}{ll}
	p_i & \text{if } x_i=1,\\
	1-p_i & \text{if } x_i=0.
	\end{array}
	\right.	\label{eq:x_foc}
\end{align}

Then, within model \eqref{eq:y_foc}-\eqref{eq:x_foc}, we solve the marginalized MAP estimation:
\begin{align}
\hat{\xv} = \argmax_\xv p(\xv|\yv),\label{eq:foc_VBEM}\\
\text{with}\quad\quad p(\xv|\yv) =  \int_\tv p(\xv,\tv|\yv),
\end{align}
and resort - following the comparison exposed in subsection \ref{subsec:calib} in the Gaussian case - to a Bayesian Mean-Field  approximation. The particularization of the algorithm to the Bernoulli model \eqref{eq:x_foc} is detailed in the appendix, an implementation is also available on author's webpage.

\subsection{Experiments and results}
\begin{figure}[h!]
\begin{center}
	\includegraphics[width=0.58\columnwidth]{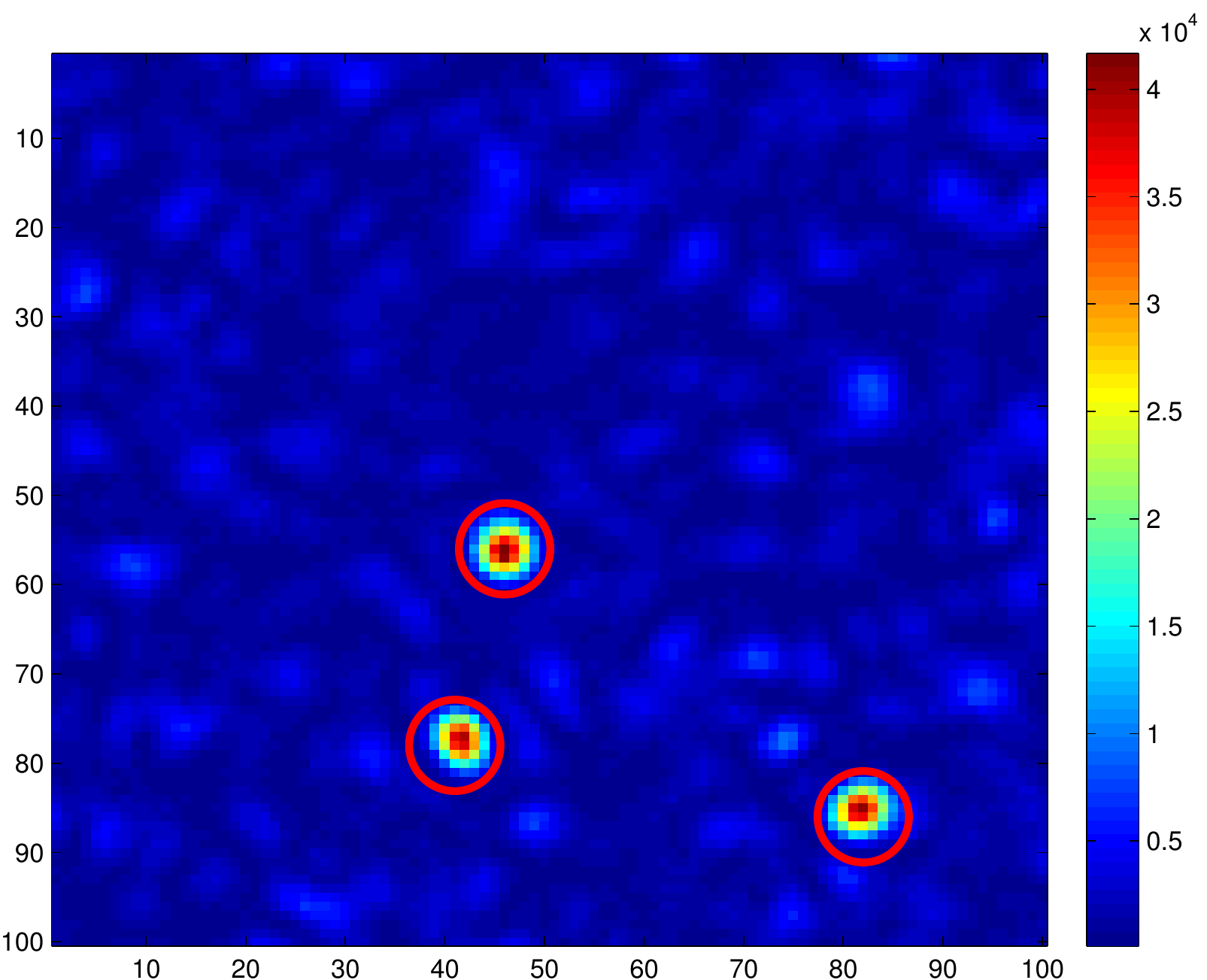}
\caption{Illustration of light focusing on 3 points. The circles mark the positions of the targets.}\label{fig:illufoc}
\end{center}
\end{figure}	

\begin{figure}[h!]
\begin{center}
	\includegraphics[width=0.58\columnwidth]{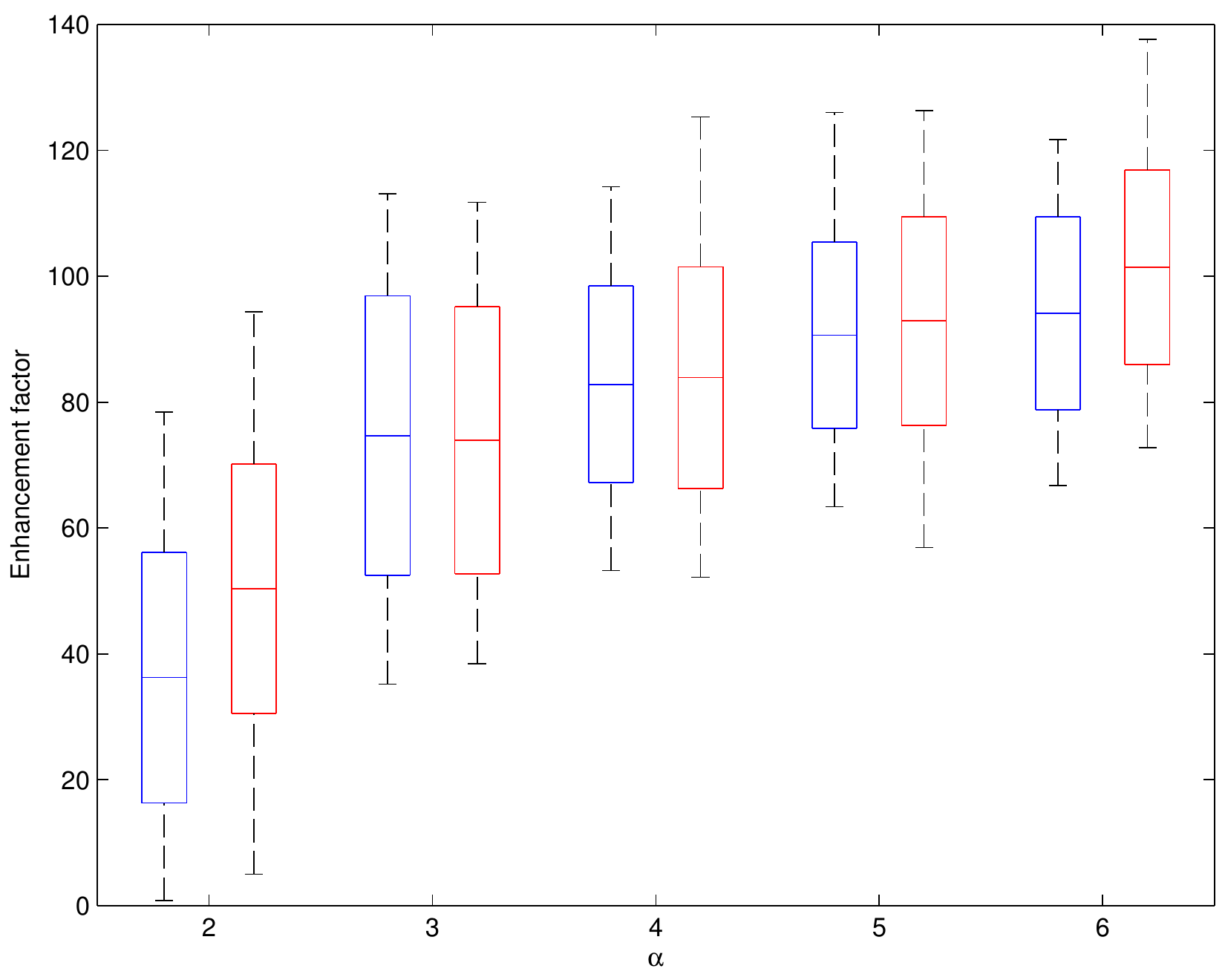}
\caption{Single target experiment. Enhancement factor as a function of the number of measurements used to learn the TM (x-axis is $\alpha$, such that $p=\alpha N$ calibration measurements are used). For the same estimation of the TM, 2 focusing techniques are compared: phase conjugation (blue boxes), and the new Mean-Field technique (red boxes).}\label{fig:illumoustache}
\end{center}
\end{figure}

In this section, we  assess the performance of the proposed focusing approach through different experiments. The general setting is as follows. The DMD inputs, here taken of dimension $N=1600$, are estimated according to the procedure described above from the desired outputs, focusing on $1$ to $4$ target points. We set the Bernoulli parameters $p_i$ to $0.5$, noticing that asymptotically half of the DMD pixels  are expected to be ``ON'' (\cite{akbulut_focusing_2011}).
Finally, the TM, reduced to its rows of interest, is measured as discussed in section \ref{sec:pr}. 

Fig. \ref{fig:illufoc} shows an example of the observed output field, corresponding to the estimated  DMD configuration, optimized to focus on 3 points. To quantitatively evaluate the focusing performance,   
we measure
the intensity enhancement factor, as:
\begin{align}
\eta \triangleq \frac{I_{\text{foc}}}{I_{\text{back}}},
\end{align}
where $I_{\text{foc}}$ is the intensity inside the target area after spatial binary amplitude modulation is performed, $I_{\text{back}}$ is the average background intensity. 
This value is measured for $100$ trials, as a function of the number of calibration measurements used to learn the TM.
Two different setups are then considered: the single-point focusing case and the multi-target case.

\subsection{Focusing on a single point}

Fig. \ref{fig:illumoustache} compares the enhancement factors achieved by two different focusing methods, namely a simple phase-conjugation - performing $\hat{\xv} = \Big[ \Re(\D^H\yv)>0 \Big]$ - and the proposed method, in the case where only one target point is focused. Results are presented under a ``box'' format, where:
\begin{itemize}
\item the middle segment stands for the average enhancement $\bar{\eta}$ over the $100$ trials, 
\item the upper and lower bounds of the rectangle define the interval $[\bar{\eta}-\sigma_\eta\quad \bar{\eta}+\sigma_\eta]$ (where $\sigma_\eta$ is the experimentally computed standard deviation), in which lies, under the Gaussian assumption, 68 \% of the trials,
\item the whiskers represent the minimum and maximum values observed over the entire set of trials.
\end{itemize}
For each experiment point $\alpha\in\lbrace2,\ldots,6\rbrace$, such that $p=\alpha N$ calibration measurements are used to compute the TM, we display the boxes related to the phase-conjugation method (blue boxes), and the new Mean-Field technique (red boxes) described in the previous section. 

As a first observation, we can see that the general dependency with regard to $\alpha$ noticeably resonates with the curve of the \textit{prVBEM} algorithm in Fig. \ref{fig:corr}: there is a clear gap between the performance achieved for $\alpha=2$ and for $\alpha=3$, while, for $\alpha\geq3$, the intensity enhancement keeps increasing but less significantly. 

Interestingly, the Mean-Field approach seems to outperform phase-conjugation, with regard to the mean and maximum values measured, but not always in a statistically significant manner. Focusing on the most favorable case considered here, namely with $\alpha=6$, the best intensity enhancement  factor lies around $140$, to be compared with the ideal expected enhancement given by $1+\frac{1}{\pi}\left(\frac{N}{2}-1\right) \simeq 255$, see \cite{akbulut_focusing_2011}. 

\subsection{Focusing on multiple points}
For this second setup, we are interested in the performance of the proposed algorithm in a context of multiple target points. Additionally to the intensity enhancement, we consider here the missed detection rate, defined as the number of trials (expressed in percentage) failing to focus on \textit{at least} one of the multiple target points, \ie the number of trials for which at least one of the $T$ largest intensity peaks in the output image does not match any of the $T$ targets.

\begin{figure}[h!]
\begin{center}
	\captionsetup[subfigure]{labelformat=empty}
	\subfloat[(a) Enhancement factor]{\includegraphics[width=0.48\columnwidth]{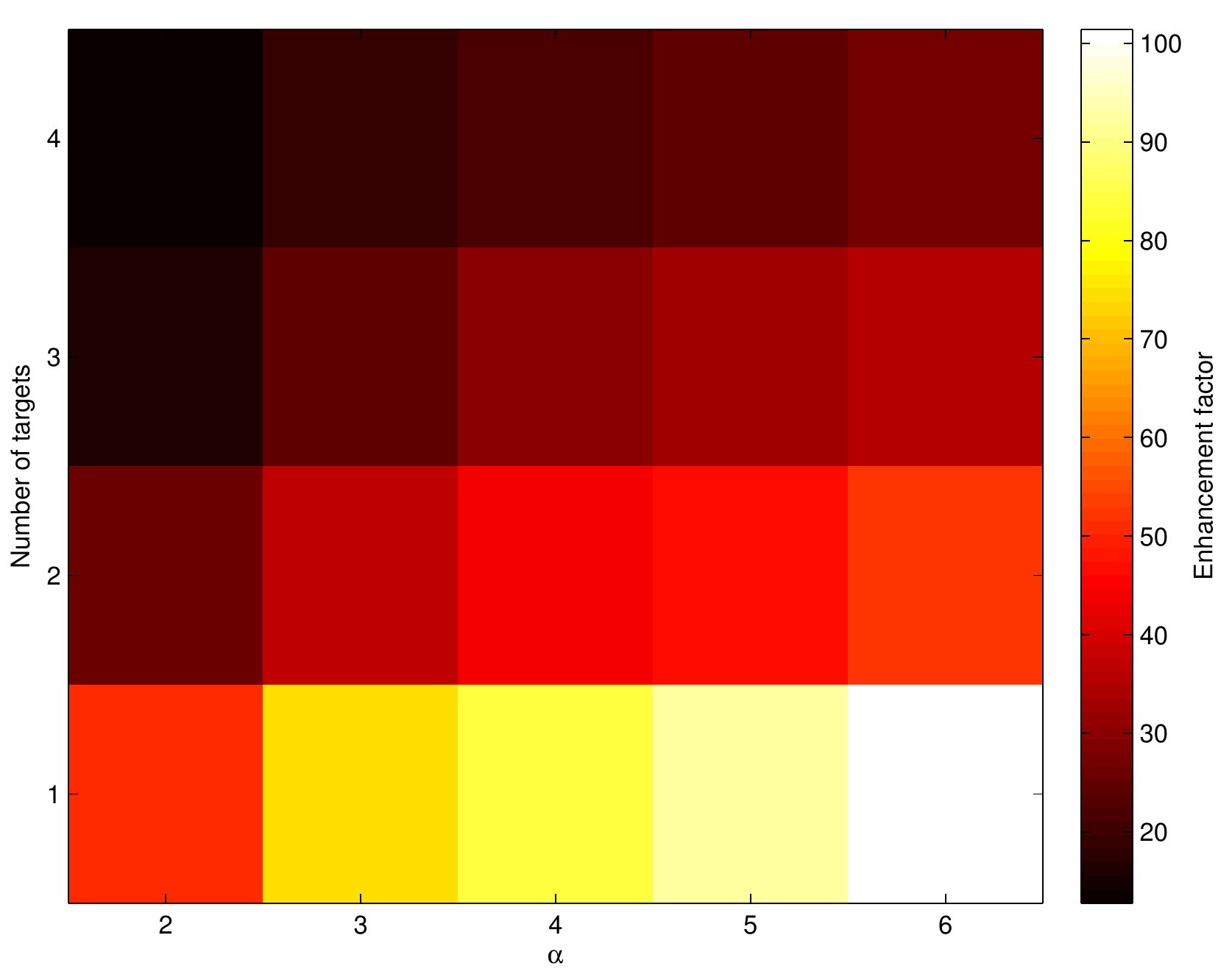}\label{fig:enh}}\hspace{0.4cm}
	\subfloat[(b) Missed detection rate]{\includegraphics[width=0.48\columnwidth]{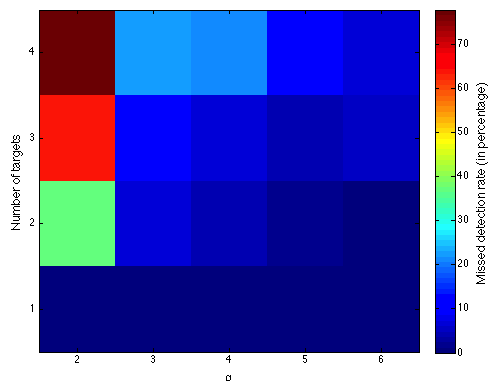}\label{fig:mdr}}
	\caption{ Multiple target experiment. (a) Average enhancement factor as a function of the number of measurements used to learn the TM (x-axis is $\alpha$, such that $p=\alpha N$ calibration measurements are used), and the number of target points (y-axis). (b) Missed detection rate   (same axis as in (a)).}\label{fig:multitargets}
\end{center}
\end{figure}

Fig. \ref{fig:multitargets} represents these two figures of merit under diagram formats. They present an interesting general symmetry: increasing the number of targets or decreasing the number of calibration points leads to an increase of the missed detections and a decrease of the enhancement factor. The missed detection rate seems however more sensitive to the number of calibration points used to learn the TM: for $\alpha=2$ and $2$ target points, the algorithm fails with a rate approaching 40\%, while for $\alpha=3$ and the same number of targets, we keep a reasonable performance (around $10$\%). In a more general view, these figures greatly highlight the deep relation between the quality of the calibration and the focusing performance.

\section{Conclusion}
This paper shows that the full \textit{complex-valued} transmission matrix of a strongly scattering material can be estimated, up to a global phase factor on each of its rows, with a simple experimental setup involving only \textit{real-valued} inputs and outputs. In our experiment, the  inputs are amplitude modulations on a \textit{binary}  DMD, and the output is the field intensity measured on a CCD camera, that gathers a significant amount of measurement noise. Note that no reference arm is used, that would allow interferometric measurements, but that would make the experimental setup  more complex and considerably more unstable. 

We here resort to Bayesian phase retrieval techniques, and we have shown that, amongst such techniques,  a recently proposed  variational approach (VBEM) \cite{Dremeau} allows a precise estimation of the transmission matrix, tractable in computational complexity and scalable for large-size signals, provided that we have a sufficiently large number of input-output calibration signals. 
Experimental  results validate this concept, both in terms of output prediction, distribution of singular values, and  in an application of light focusing  onto a number of target points in the output plane. 
It should be emphasized that this estimation of the transmission matrix opens many applications beyond light focusing, may it be for imaging through the scattering material \cite{popoffNJP2011, Liutkus_ScientificReports2014}, or for obtaining information about the scattering material itself. 

\section{Appendix: Focusing with a Mean-Field based algorithm\label{sec:app}}
The VBEM algorithm is an iterative procedure which successively updates the factors of the Mean-Field approximation. Particularized to model  \eqref{eq:y_foc}-\eqref{eq:x_foc}, this gives raise to the following update equations:
\begin{align}
q(\theta_\mu) &=  \frac{1}{2\pi\; I_0(\frac{2}{\sigma^2}|y_\mu^*\langle z_\mu\rangle|)}\;\exp\left(\frac{2}{\sigma^2} \Re(y_\mu^*\langle z_\mu\rangle e^{j\theta_\mu})\right),\\
q(x_i) &= p(x_i)\; \exp\left(x_i \;\frac{2\;\Re(\dv_i^H\langle \rv_i\rangle)-\dv_i^H\dv_i}{\sigma^2}\right),
\end{align}
where
\begin{align}
&\langle \rv_i \rangle=\bar{\yv}-\sum_{k\neq i} q(x_k=1)\;\dv_k,\label{eq:resi2}\\
&\bar{\yv}=\left [y_\mu e^{(j \arg(y_\mu^*\langle z_\mu\rangle))}\;\frac{I_1(\frac{2}{\sigma^2}|y_\mu^*\langle z_\mu\rangle|)}{I_0(\frac{2}{\sigma^2}|y_\mu^*\langle z_\mu\rangle|)}\right]_{\mu=\lbrace1\ldots N\rbrace},\\
& \langle z_\mu\rangle = \sum_i q(x_i=1)\; d_{\mu i},
\end{align}
and $I_0$ (resp. $I_1$) stands for the modified Bessel function of the first kind for order $0$ (resp. $1$).

Coming back to problem \eqref{eq:foc_VBEM}, an approximation of $p(\xv|\yv)$ thus simply follows from
\begin{align}
p(\xv|\yv)
& = \int_\tv p(\xv,\tv|\yv),\\
& \simeq \int_\tv\prod_i q(x_i)\prod_\mu q(\theta_\mu),\\
& = \prod_i q(x_i).
\end{align}
Using this approximation, the problem is then easy to solve by a simple thresholding operation, \ie $\hat{x}_i=1$ if $q(x_i=1)>0.5$ and $\hat{x}_i=0$ otherwise.

\section*{Acknowledgements}
AD is currently working at ENSTA Bretagne, STIC/AP, 2 rue François Verny, F-29200 Brest, France. OK acknowledges the support of the Marie Curie Intra-European Fellowship for career development (IEF). LD acknowledges a joint research position with the Institut Universitaire de France. This work has been supported in part by the CSI:PSL grant and LABEX WIFI (Laboratory of Excellence within the French Program ``Investments for the Future'') under references ANR-10-LABX-24 and ANR-10-IDEX-0001-02 PSL*, and by the ERC under the European Union's 7th Framework Programme Grant Agreements 307087-SPARCS and 278025-COMEDIA.
\end{document}